# Néel-type optical skyrmions inherited from evanescent electromagnetic fields with rotational symmetry

Bo Tian[1,‡], Jingyao Jiang[1,‡], Ningsheng Xu[1,2], Zebo Zheng[1], Ximiao Wang[1], Shaojing Liu[1], Wuchao Huang[1], Tian Jiang[3,*], Huanjun Chen[1,*], Shaozhi Deng[1,*]

[1]State Key Laboratory of Optoelectronic Materials and Technologies, Guangdong Province Key Laboratory of Display Material and Technology, School of Electronics and Information Technology, Sun Yat-sen University, Guangzhou 510275, China.
[2]The Frontier Institute of Chip and System, Fudan University, Shanghai 200433, China.
[3]Institute for Quantum Information Science and Technology, College of Science, National University of Defense Technology, Changsha 410073, China.

**KEYWORDS.** optical skyrmions, surface plasmon polaritons, evanescent electromagnetic fields, rotational symmetry, two-dimensional atomic crystals

**ABSTRACT.** Optical skyrmions, the optical analogue of topological configurations formed by three-dimensional vector fields covering the whole $4\pi$ solid angle but confined in a two-dimensional (2D) domain, have recently attracted growing interest due to their potential applications in high-density data transfer, storage, and processing. While the optical skyrmions have been successfully demonstrated using different field vectors in both of free-space propagating and near-field evanescent electromagnetic fields, the study on generation and control of the optical skyrmions, and their general correlation with the electromagnetic (EM) fields, are still in infancy. Here, we theoretically propose that an evanescent transverse-magnetic-polarized (TM-polarized) EM fields with rotational symmetry are actually Néel-type optical skyrmions of the electric field vectors. Such optical skyrmions maintain the rotation symmetry that are independent on the operation frequency and medium. Our proposal was verified by numerical simulations and real-space nano-imaging experiments performed on a graphene monolayer. Such a discovery can therefore not only further our understanding on the formation mechanisms of EM topological textures, but also provide a guideline for facile construction of EM skyrmions that may impact future information technologies.

The generation of vortex textures and study of their dynamics have been a hot topic in various research fields covering high-energy physics,[1,2] materials science,[3,4] condensed matter physics,[5–9] and optics.[10,11] Skyrmion is a prominent vortex configuration that is formed by topologically protected vector fields spreading in 2D domains but wrapping a whole Poincaré polarization vector sphere.[12] Because the skyrmions are topologically stable even at the nanoscale,[13-15] they have opened new avenues for high-density information processing, transfer, and storage.

Previously the skyrmions have been widely demonstrated and studied in magnetic materials,[16] Bose−Einstein condensates,[17] liquid crystals,[18] as well as quantum Hall system.[19] In recent years, optical skyrmions have been attracting increasing research interest because of their promising applications in high-density optical data storage and processing.[12] In 2018, optical skyrmion lattice of electric fields was for the first time demonstrated by interferences of surface plasmon polaritons in a metal film[12]. Afterwards, various types of optical skyrmions have been realized and studied, although most of them are from theoretical aspect, using electric fields,[12,20–25] stokes vector field,[26] magnetic fields,[23,28] or photonic spin angular momentum,[29-37] either in free-space propagating EM waves or in confined evanescent waves. These excellent studies unveil the new degrees of freedom introduced by the optical skyrmions for tailoring light−matter interactions at sub-wavelength scale, which might enable a variety of applications in advanced optical microscopy and ultra-compact photonic devices.

Despite the aforementioned progresses, a facile manner for generating optical skyrmions without introducing complex and stringent excitation configuration is still challenging. Moreover, the general correlation of the skyrmions with the EM fields that generate them still needs an in-depth study. Here, by analyzing the general mathematical form of the EM field, we find that the TM-polarized evanescent fields with rotational symmetry is actually a Néel-type optical skyrmion of the electric field vectors. In addition, skyrmions with topological invariant $S$ ($S$: also known as skyrmion number) of 1 (skyrmion) and −1 (anti-skyrmions) can be observed alternatively in the concentric rings extending outward from the center of the evanescent fields. The theoretical results are consistent with numerical simulations, and are further verified experimentally using surface plasmon polaritons (SPPs) supported in a monolayer graphene. These topological textures are protected by the rotation symmetry and are independent on the excitation frequency and working medium, which can in principle be extended to any generic system supporting TM evanescent fields with rotational symmetry. We therefore believe that these results can not only provide a facile guideline for generating optical skyrmions, but also further our understanding on the EM topological textures.

A skyrmion is a topological-protected structure consisting of a three-dimensional (3D) vector field (spin, magnetic field, electric field, stokes vector field, *etc.*) distributing in a 2D domain. In principle there are two criteria to determine whether an architecture is a skyrmion or not. One is to observe that the vector field smoothly evolves and covers the entire Poincaré polarization vector sphere, the other is to calculate the topological invariant of the system. Specifically, the topological invariant $S$ is defined in Cartesian coordinate system as:

$$S = \frac{1}{4\pi}\iint_D s\,dxdy = \frac{1}{4\pi}\iint_D \vec{e}\cdot(\partial_x\vec{e}\times\partial_y\vec{e})\,dxdy \qquad (1)$$

where $s = \vec{e}\cdot(\partial_x\vec{e}\times\partial_y\vec{e})$ is the topological invariant density, $D$ is the integration area, and $\vec{e}$ is the normalized 3D vector field. When the $S$ calculated by Equation (1) is $\pm 1$, it means that the direction of the vector distributing in the region $D$ covers the whole Poincaré sphere, and that the vector field is a skyrmion[12].

In a Néel-type skyrmion the vector field has a zero angular component $e_\theta$ (Figure 1a), making the $\vec{e}$ only has its radial component $e_r$ and $z$-component $e_z$. In addition, the $\vec{e}$ should progressively changes from "up-state" in the center to "down-state" along the radial direction (Figure 1a, lower panel). Accordingly, the vector $\vec{e}$ can be expressed as $\vec{e} = \cos\varphi\,\hat{i}_r + \sin\varphi\,\hat{i}_z$. The angle $\varphi$ is defined as the angle between $\vec{e}$ and 2D plane supporting the skyrmion (Figure 1a, lower panel, more details on definition of $\varphi$ can be found in Note S1, Supporting Information), which is a function of the coordinate $r$ and expressed as:

$$\varphi = \begin{cases} \arctan(e_z/e_r), & e_r \geq 0 \\ -\arctan(e_z/e_r), & e_r < 0 \end{cases} \qquad (2)$$

Equation (1) can then be simplified to a one-dimensional integral $S = -\frac{1}{2}\int_D \cos\varphi\frac{\partial\varphi}{\partial r}dr = \int_D n_s dr$, with the definition of $n_s$ as $n_s = -\frac{1}{2}\cos\varphi\frac{\partial\varphi}{\partial r}$ (more mathematical details can be found in Note S2, Supporting Information). It is noted that for this one-dimensional integral, $S$ is always equal to 1 (−1) as long as the angle $\varphi$ changes from $\pi/2$ (−$\pi/2$) to $-\pi/2$ ($\pi/2$). According to these analysis, one can generate Néel-type optical skyrmions by constructing EM fields fulfilling the two criteria: i) they are rotationally symmetric without the angular components, and ii) the field vectors transform radically from the up-state to the down-state.

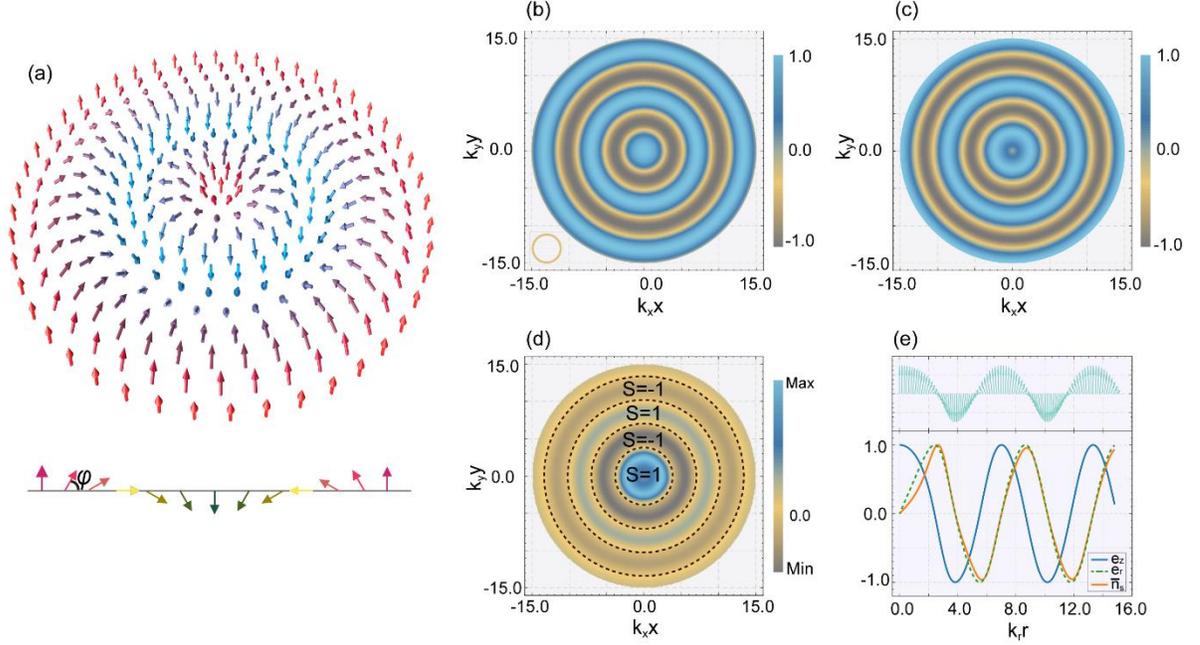

**Figure 1.** Néel-type skyrmions. (a) Distribution of the vector fields in a Néel-type skyrmion with rotational symmetry. Lower panel: distribution of the vector fields along a specific radial direction, showing a cycling from the up-state ($\varphi = 90°$) at the skyrmion center to down-state ($\varphi = -90°$) and then back to up-state again. (b−d) Distributions of the normalized $E_z$ ($e_z = E_z/|\vec{E}|$) (b), normalized $E_r$ ($e_r = E_r/|\vec{E}|$) (c), and topological invariant density $s$ (d) in the $x-y$ plane. Bottom left in (b): schematic showing a circle in the momentum space. The boundaries of a specific skyrmion with a topological invariant $S$ equaling 1 or −1 are marked by the two adjacent dashed lines shown in (d). (e) The radial variations of the $e_z$, $e_r$, and normalized $n_s$ (defined as $\bar{n}_s = n_s/n_{s,\max}$) along a specific

in-plane direction, which are respectively extracted from (b−d). Upper panel: radial variation of the electric field vector. The electric fields in (b), (c), and (e) belong to a TM-polarized evanescent EM field, whose $E_z$ component is generated by the inversed Fourier transformation of the momentum circle shown in the bottom left of (b). The topological invariant density shown in (d) is calculated with the electric fields corresponding to (b) and (c).

We propose that a TM-polarized evanescent EM field, which exhibits rotational symmetry with zero tangential component, is a Née-type optical skyrmion of the electric fields. Specifically, we consider an evanescent field decaying along the $z$-axis, which should depend on $z$ in the form of $e^{-k_z z}$. In a cylindrical coordinate system with $(r, \theta, z)$, the relation between the different electric field components $(E_r, E_\theta, E_z)$ of the TM-polarized field can be expressed as (Note S3, Supporting Information):

$$\begin{cases} E_r = -\dfrac{k_z}{k_s^2} \dfrac{\partial E_z}{\partial r} \\ E_\theta = -\dfrac{k_z}{r k_s^2} \dfrac{\partial E_z}{\partial \theta} \end{cases} \tag{3}$$

where $k_s$ is the in-plane wave vector component. Due to the rotational symmetry, the $E_z(r, \theta) = AI(r)$ and $E_\theta$ is always 0, with $A$ representing the amplitude of the electric field. The electric field can then be expressed as $\vec{E} = E_r \hat{i}_r + E_z \hat{i}_z$. The angle $\varphi$ corresponding to the angle between the electric field vector and the plane $z = 0$ can be determined according to Equation (2). Consequently, it can be clearly seen that $\varphi$ is $\pi/2$ or $-\pi/2$ when $E_r$ is 0. For a continuous time-harmonic EM waves, the transition of the angle $\varphi$ from $\pi/2$ to $-\pi/2$, *i.e.*, the field vector transform from up-state to down-state, can always be satisfied, thus consisting an optical skyrmion as shown in Figure 1a.

We then verify the above theoretical result with a simple example, which can also be realized experimentally as discussed below. Mathematically, a rotationally symmetric TM field with zero tangential component can be synthesized by performing the inverse Fourier transform (FT) on a function $F(k_x, k_y)$ of the same rotational symmetry (Note S4, Supporting Information). The simplest form of $F(k_x, k_y)$ is a circle (Figure 1b, lower left corner), where its radius represents the mode of the isotropic wave vector. If we set the radius of the momentum circle as $k_s$, then $F(k_x, k_y) = C\delta(\rho - k_s)$, with $C$ a constant and $\rho$ the radial coordinate in the momentum space. The inverse FT of such a circle then corresponds to:

$$f(r) = C k_s J_0(k_s r) \tag{4}$$

where $J_0$ is a zero-order Bessel function in the real space (Note S4, Supporting Information). If we set $E_z = f(r)$, the whole electric field distribution in the real space can be obtained using Equation (3) and (4). The normalized in-plane component $e_r$ ($e_r = E_r/|\vec{E}|$) and z-component $e_z$ ($e_z = E_z/|\vec{E}|$) of the electric field are shown in Figure 1c and 1b, respectively, which are of rotational symmetry as expected. With the knowledge of $E_z$ and $E_r$ the topological invariant density $s$ can be readily calculated using $s = \vec{e} \cdot (\partial_x \vec{e} \times \partial_y \vec{e})$, which is also rotationally symmetric as shown in Figure 1d. The radial variations of the $e_z$, $e_r$, and normalized $n_s$ (defined as $\bar{n}_s = n_s / n_{s,\max}$) along a specific in-plane direction are shown in Figure 1e. Notably, the electric field vector rotates from the up state to down sate consecutively along the radial direction (Figure 1e, upper part) and agrees well with that shown in Figure 1a (lower part), which is a typical feature of the Néel-type skyrmions. It can also be seen that the maxima/minima of $e_z$ correspond exactly to zeros of $e_r$. In particular, according to the definition of $n_s = -\frac{1}{2}\cos\varphi \frac{\partial \varphi}{\partial r}$, a characteristic of the Néel-type skyrmion we propose is that, the zero points of $n_s$ and $e_r$ ($\varphi = \pm\pi/2$), always coincide with each other. Such a result clearly indicates that the $n_s$ is 0 when the electric field vector is perpendicular to the $x-y$ plane, as manifested in Figure 1e. One can readily calculate the topological invariant $S$ according to Equation (1). When the integral interval of Equation (1) is defined as the coordinate range corresponding to any two adjacent zeros of $e_r$ (*e.g.*, the two adjacent dashed lines shown in Figure 1d), the obtained $S$ is 1 or −1. Therefore, a series of optical skyrmions are generated in the form of electric field vectors (Figure 1d, the ring regions separated by two adjacent dashed lines).

According to the above analysis, EM skyrmions can be generated by introducing a rotationally symmetric TM-polarized evanescent wave. SPPs are electromagnetic modes formed by coupling of collective free electron oscillations and photons. They only exist in TM polarization and are evanescent in nature, which exponentially decay in the direction perpendicular to the interface between a conductor and a dielectric.[38] Thus, SPPs waves which propagate isotropically in the $x-y$ 2D plane fulfill the requirement for formation of the Néel-type skyrmions as proposed. Among the various plasmonic materials, monolayer graphene is an outstanding platform supporting TM-polarized SPPs waves covering a broad spectral range from the ultraviolet to millimeter wave regime.[39] Due to the isotropic in-plane dielectric function of the monolayer graphene, the SPPs waves will propagate isotropically within the graphene plane.[38] In particular, the monolayer graphene is able to confine the EM fields down to the deep-subwavelength scale,[39] whereby optical skyrmions with much smaller sizes than those supported by other plasmonic materials (*e.g.* metal) can be expected. All these features make the monolayer graphene an ideal physical platform for creating and studying the optical skyrmions.

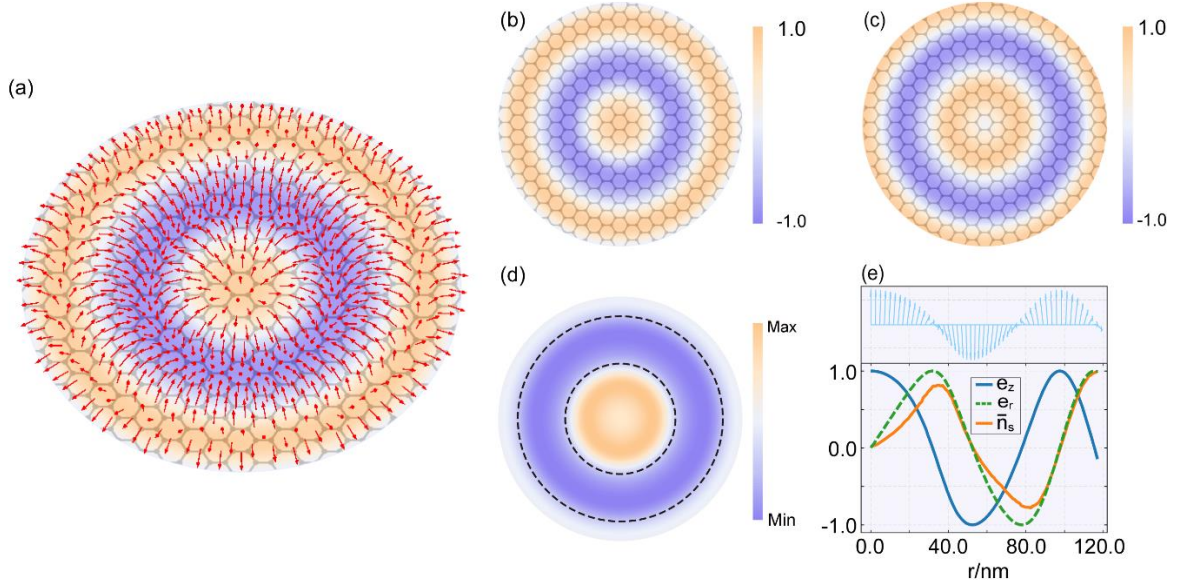

**Figure 2.** Néel-type optical skyrmions formed by surface plasmon polaritons in a monolayer graphene. (a) Spatial distribution of the three-dimensional normalized electric field vectors, $\vec{E}/|\vec{E}|$, within the monolayer graphene plane. (b–d) Distributions of the normalized $E_z$ ($e_z = E_z/|\vec{E}|$) (b), normalized $E_r$ ($e_r = E_r/|\vec{E}|$) (c), and topological invariant density $s$ (d) in the monolayer graphene. Boundaries of a skyrmion with topological invariant $S$ of 0.995 (≈ 1) and −0.994 (≈ −1) are marked by the two adjacent dashed lines in (d). (e) The radial variations of the $e_z$, $e_r$, and normalized $n_s$ (defined as $\bar{n}_s = n_s/n_{s,\max}$) along a specific in-plane direction, which are respectively extracted from (b–d). Upper panel: radial evolvement of the electric field vector. The topological invariant density shown in (d) is calculated using the electric fields corresponding to (b) and (c). The excitation wavelength is 10.70 μm.

To demonstrate the formation of EM skyrmions, we employed a *z*-polarized electric dipole to excite SPPs waves at the interface of air and a monolayer graphene (see Methods part for simulation details), resulting in an isotropically propagating evanescent waves (Figure 2a). The spatial distribution of the normalized 3D electric field vectors, $\vec{E}/|\vec{E}|$, within the monolayer graphene plane clearly manifests a rotational symmetry, where the field vectors rotate continuously and alternatively between up-state to down-state along the radial direction (Figure 2a and Figure 2e, upper panel). In addition, the $e_z$, $e_r$, and $s$ all exhibit similar rotational symmetry (Figure 2b−2d). These characteristics are exactly the same as those shown in Figure 1, indicating that optical skyrmions of electric field can be created in a monolayer graphene by excitation of the isotropically propagating SPPs waves. The periodic oscillations of the $e_z$, $e_r$, and $\bar{n}_s$ along a specific radial direction can be observed in Figure 2e, which indicate that when the $e_r$ equals zero, the $\bar{n}_s$ and $|e_z|$ become zero and maximum (Table S1, Supporting Information), respectively. These features are also consistent with those predicted by Figure 1e. The integration of $n_s$, *i.e.*, the topological invariant $S$, is 0.995 or −0.994 (Table S2, Supporting Information), which are very close to 1 and −1, respectively. There are two main reasons why the topological invariants are not strictly equal to ±1. The first is the rounding error in the numerical calculation process, and the second is that the upper and lower bounds of the integral cannot be strictly located in the upper and lower bounds of the theoretical integral. Therefore, Néel-type optical skyrmions are established in the monolayer graphene, as demonstrated by the regions enclosed by two adjacent dashed lines shown in Figure 2d.

We then experimentally constructed and demonstrated the optical skyrmions of electric fields using a monolayer graphene. To that end, the graphene is structured into a circular disk with a diameter of 500 nm (Figure 3a). Nano-imaging technique based on a scattering-type scanning near-field optical microscope (s-SNMO: NeaSNOM, Neaspec GmbH, Munich, Germany) was employed to inspect the localized electric field distributions within the graphene disk. It should be noted that it is crucial to have the nano-imaging technique capable of resolving the phase of $E_z$ within the graphene disk, otherwise one cannot extract the full electric field vectors from the recorded near-field distributions and calculate the topological invariant (more details on calculating the different electric field components from the experimental near-field image are provided in Note S5, Supporting Information). SPPs waves were launched into the graphene plane by focusing a mid-infrared laser ($\lambda$ = 10.70 μm) onto a metallic nanotip, which was placed 60 nm above the circular disk (see Methods part for more details) (Figure 3a). By scanning the graphene disk underneath the metallic tip and collecting the backscattering light from the tip, the electric fields can be probed and analyzed. Upon excitation, the polariton waves propagated isotropically within the disk and were reflected by the boundary. Because the circular disk is rotationally symmetric, the reflected waves from different positions on the boundary will propagate and interfere with those launched from the tip, forming a standing waves of the same rotational symmetry.[35] In this way, a physical system capable of supporting the optical skyrmions was realized experimentally.

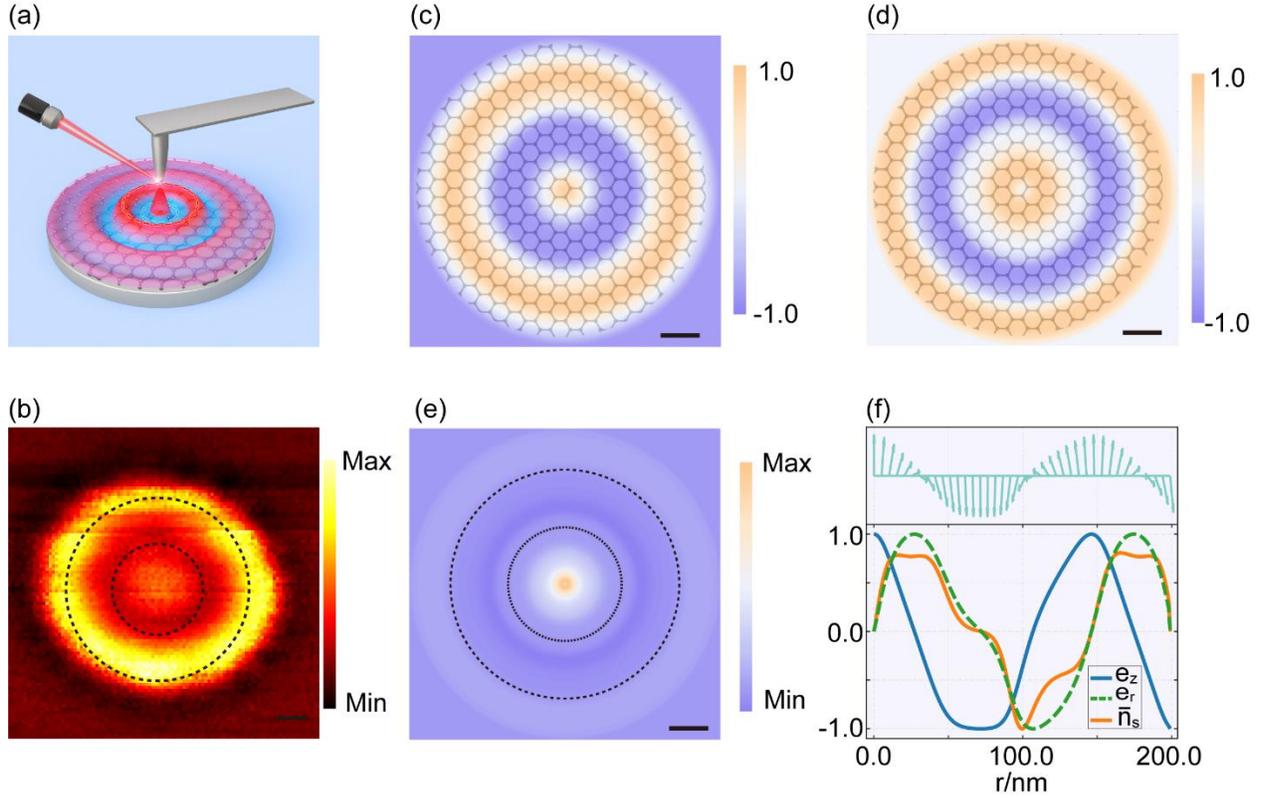

**Figure 3.** Experimental realization of the Néel-type optical skyrmions. (a) Schematic showing construction and observation of the Néel-type optical skyrmions in a graphene circular disk. (b) Experimental near-field distribution of $|E_z|$ within a graphene disk. (c, d) Reconstructed $e_z$ (c) and $e_r$ (d) within the graphene disk. (e) Topological invariant density $s$ in the graphene disk. Boundaries of a skyrmion with topological invariant $S$ of 1.007 ($\approx 1$) and $-1.001$ ($\approx -1$) are marked by the two adjacent dashed lines. (f) The radial variations of the $e_z$, $e_r$, and $\bar{n}_s$ along a specific radial direction, which are respectively extracted from (c−e). Upper panel: radial variation of the electric field vector. The topological invariant density shown in (e) is calculated with the electric fields corresponding to (c) and (d). The excitation wavelength is 10.70 μm.

The SPPs interferences create a near-field distribution of the $|E_z|$ in the graphene disk, showing well-defined ring-shaped patterns with evident rotational symmetry (Figure 3b).

Furthermore, the phase image also presents the same symmetry as the $|E_z|$ exhibits (Figure S2, Supporting Information), where the phase transforms continuously from $0.09\pi$ at the center to $0.17\pi$ at the first brightest fringe and then to $-0.06\pi$ at the darkest fringe along the radial direction (Figure S2, Supporting Information). The $e_z$ can then be readily extracted according to the phase distribution (Note S5, Supporting Information) (Figure 3c), whereupon the $e_r$ can be obtained according to Equation (3) (Figure 3d). Although these two electric field components are a bit different from the theoretical (Figure 1b and 1c) and simulation (Figure 2b and 2c) results, they indeed resemble the electric field configurations shown in Figure 1 and 2. Additionally, both of them rotate radially (Figure 3f, lower panel), making the resulting total electric field vector varying cyclically between $\pi/2$ (up-state) and $-\pi/2$ (down-state), as expected (Figure 3f, upper panel). The $\bar{n}_s$, which is extracted from the topological invariant density $s$ (Figure 3e), follows the same trend as that predicted in Figures 1e and 2e and quantitatively, has the same zero coordinates as shown by the $e_r$. The calculated $S$ are 1.007 and $-1.001$ in the two regions enclosed by the dashed lines (Figure 3e), respectively, thus demonstrating the formation of two optical skyrmions.

The above analyses well demonstrate that the Néel-type optical skyrmion is actually an intrinsic characteristic of a TM-polarized evanescent EM field with rotational symmetry. This feature is independent either on the operation frequency or the medium that supports the EM field. We employed numerical simulations to show this more clearly. If we change the excitation frequency to 30 THz (10.00 μm), similar optical skyrmions can be observed on the graphene disk (Figure S3, Supporting Information). The optical skyrmions are shrunk due to the shorter polariton wavelength. The $z$-polarized electric dipole was further utilized to excite a gold flake (Figure S4, Supporting Information) and a hexagonal boron nitride (hBN) flake (Figure S5, Supporting Information), which respectively support TM-polarized SPPs and phonon polaritons (PhPs) in visible and mid-infrared spectral range. For the gold flake, the excitation frequencies were 564 THz ($\lambda$ = 532 nm) and 474 THz ($\lambda$ = 633 nm), while for the hBN flake, excitation frequencies of 24 THz ($\lambda$ = 12.50 μm) and 23.68 THz ($\lambda$ = 12.66 μm) were employed. Both of the gold and hBN flakes exhibit isotropic in-plane permittivities, whereby the polariton waves excited by the dipole propagate isotropically and form evanescent TM fields with rotational symmetry. The Néel-type optical skyrmions of the electric fields are then evidently observed in both of the gold (Figure S4, Supporting Information) and hBN flakes (Figure S5, Supporting Information).

It is worth noting that in the numerical simulations (Figure 2b and 2c) and experimental measurements (Figure 3b−3d), the polariton fields generated in monolayer graphene are physically different from each other because one of them is launched directly by the electric dipoles (numerical simulation), while the other is formed by the interference between tip-launched polariton waves and those reflected by the boundaries (experimental measurement). This results in discrepancies between the calculated (Figure 2e) and measured (Figure 3f) $e_z$, $e_r$, and especially, the $\bar{n}_s$. However, according to our theoretical study, both of the two fields give rise to Néel-type optical skyrmions of the electric fields, because they both exhibit rotational symmetry, TM-polarization, and are of evanescent nature. This is a direct manifestation of the topological robustness of the skyrmions. Following this analysis, as long as an evanescent field is rotationally symmetric and TM-polarized, it must give optical skyrmions, regardless of how the field is generated.

The optical skyrmions formed in the graphene circular disk can have additional merits in terms of the skyrmion size and tunability. First, the atomic-scale thickness of the monolayer graphene and its SPPs with ultra-strong EM field localization can endow the optical skyrmions with sizes much smaller than the diffraction limit of electromagnetic waves. For example, in our nano-imaging study, the lateral distribution of skyrmions is only on the order of ~100 nm, while the vertical distribution is only on the order of ~10 nm. These sizes are about 1/100 of the excitation wavelength ($\lambda$ = 10.70 μm), which already penetrate into the deep sub-wavelength region. Secondly, the graphene electronic structure and therefore the permittivity can be facilely tailored by external stimuli, making it possible to create skyrmions that are tunable through electrical gating, optical pumping, chemical doping, or thermal stimulation. All these merits make skyrmions formed in monolayer graphene have important application potentials in high-speed optical signal processing and large-capacity information storage.

In summary, we propose, simulate, experimentally verify the existence of Néel-type optical skyrmions in TM-polarized evanescent EM fields. Such skyrmions are in the form of electric field vectors and are irrespective of the operation wavelength and medium, whose sizes are in the deep subwavelength spatial regions if the evanescent fields are formed by polariton waves. The results offer a facile method to construct optical skyrmions, which may greatly benefit future information devices operating in a broad spectral ranges, processing at a high speed, as well as with an ultra-compact volume.

**Methods**

**Sample fabrication and measurements**

Monolayer graphene was transferred onto a silicon substrate and then etched into a 500-nm-diameter circular disk using electron beam lithography and oxygen plasma etching. Optical nano-imaging was performed using a scattering-type s-SNOM (NeaSNOM, Neaspec GmbH, Munich, Germany). In a specific measurement, a metal-coated tip (Arrow-IrPt, NanoWorld) was illuminated by a mid-infrared laser (Daylight Solutions) with tunable frequency from 900 cm$^{-1}$ to 1240 cm$^{-1}$. The tip vibrated vertically with a frequency of 260 kHz. The back-scattered light from the tip interfered with the reference beam and was guided to an MCT detector (HgCdTe, Kolmar Technologies). The near-field signal was extracted by demodulating the detected signal at a fourth harmonic.

**Numerical simulations**

All simulations were performed using finite element method (Comsol Multiphysics). Specifically, SPPs on the surfaces of a monolayer graphene and a gold film, and PhPs on the surface of an hBN flake were excited by a *z*-polarized electric dipole, respectively. In each simulation, the dipole was located 100 nm above the sample surface. The near field signal, Re($E_z$), was obtained on the plane 50 nm above the sample surface. During the simulation, the conductivity of graphene is:

$$\sigma = \frac{e^2 E_f}{\pi \hbar^2} \frac{i}{\omega + i\tau^{-1}} \tag{5}$$

where $E_f$ ($E_f$ = 0.4 eV) is the Fermi energy , $\tau$ ($\tau$ = 20 fs) is the relaxation time. The thickness of the monolayer graphene is set as 1 nm. The dielectric function of gold is set according to the material library embedded in Comsol Multiphysics (RakiA et al. 1998: Lorentz-Drude model; *n*, *k* 0.248−6.20 μm). The thickness of the gold flake is 200 nm. The dielectric tensor of hBN was calculated using a Lonrentzian model:

$$\varepsilon_j = \varepsilon_\infty^j \left(1 + \frac{\omega_{j,L}^2 - \omega_{j,T}^2}{\omega_{j,L}^2 - \omega^2 - i\omega\Gamma_j}\right), j = r, z \tag{6}$$

where $\varepsilon_j$ denotes the principal components of the permittivity tensor. Parameter $\varepsilon_\infty^j$ is the high frequency dielectric constant, and $\omega_{j,L}$ and $\omega_{j,T}$ are the longitudinal and transverse optical phonon frequencies, respectively. Parameter $\Gamma_j$ is the broadening factor of the Lorentzian line shape. The values of these parameters during the simulation are $\varepsilon_\infty^r = 4.90$, $\varepsilon_\infty^z = 2.95$, $\omega_{r,L} = 1614$ cm$^{-1}$, $\omega_{z,L} = 760$ cm$^{-1}$, $\omega_{r,T} = 1360$ cm$^{-1}$, $\omega_{z,T} = 825$ cm$^{-1}$, $\Gamma_r = 10$ cm$^{-1}$, and $\Gamma_z = 18$ cm$^{-1}$, respectively. The hBN flake has a thickness of 100 nm. All of the monolayer graphene, gold film, and hBN flake were placed onto a silicon substrate with a permittivity of 11.56. The medium above them graphene is air.

ASSOCIATED CONTENT

**Supporting Information**. The Supporting Information is available free of charge at xxx. Note S1 to Note S5; Table S1, Table S2; Néel type skyrmions (Figure S1); optical skyrmions observed in a graphene circular disk (Figure S2); Néel type optical skyrmions formed by surface plasmon polaritons in a graphene circular disk (Figure S3); Néel type optical skyrmions formed by surface plasmon polaritons in a gold flake (Figure S4); Néel type optical skyrmions formed by phonon polaritons in an hBN flake (Figure S5) (PDF).

AUTHOR INFORMATION

**Corresponding Author**

*chenhj8@mail.sysu.edu.cn;

*stsdsz@mail.sysu.edu.cn;

*tjiang@nudt.edu.cn

**Author Contributions**

H.C. and S.D. conceived and initiated the study. B.T. proposed the theoretical model, performed the numerical simulations, and analyzed the data. Sample fabrication was performed by J.J., W.H., and S. L.. T.J., B.T., Z.Z., and X.W. performed the near-field optical measurements and analyzed the experimental data. H.C., N.X., T.J., and S.D. coordinated and supervised the work and discussed and interpreted the results. The manuscript was written through contributions of all authors. All authors have given approval to the final version of the manuscript.

**Funding Sources**

The authors acknowledge Dr. Runze Chen of National University of Defense Technology for valuable discussion. The authors acknowledge support from the National Natural Science Foundation of China (Grants 91963205), the National Key Basic Research Program of China (Grants 2019YFA0210200 and 2019YFA0210203), Guangdong Basic and Applied Basic Research Foundation (Grants 2019A1515011355 and 2020A1515011329). H.C. acknowledges the support from Changjiang Young Scholar Program.

**Notes**

The authors declare no competing financial interest.

and S. L.. T.J., B.T., Z.Z., and X.W. performed the near-field optical measurements and analyzed the experimental data. H.C., N.X., T.J., and S.D. coordinated and supervised the work and discussed and interpreted the results. The manuscript was written through contributions of all authors. All authors have given approval to the final version of the manuscript.

**Funding Sources**

The authors acknowledge Dr. Runze Chen of National University of Defense Technology for valuable discussion. The authors acknowledge support from the National Natural Science Foundation of China (Grants 91963205), the National Key Basic Research Program of China (Grants 2019YFA0210200 and 2019YFA0210203), Guangdong Basic and Applied Basic Research Foundation (Grants 2019A1515011355 and 2020A1515011329). H.C. acknowledges the support from Changjiang Young Scholar Program.

**Notes**

The authors declare no competing financial interest.

ABBREVIATIONS

2D, two-dimensional; EM, electromagnetic; TM, transverse magnetic; SPPs, surface plasmon polaritons; PhPs, phonon polaritons; 3D, three-dimensional; s-SNOM, scattering-type scanning near-field optical microscope; hBN, hexagonal boron nitride.

REFERENCES

(1) G. t'Hooft, Magnetic monopoles in unified gauge theories. *Nucl. Phys. B* **1974**, *79*, 276-284.

(2) Adkins, G. S.; Nappi, C. R.; Witten E. Static properties of nucleons in the skyrme model. *Nucl. Phys. A* **1983**, *228*, 552-566.

(3) McDonnell, C.; Deng, J.; Sideris, S.; Ellenbogen, T.; Li, G. Functional THz Emitters Based on Pancharatnam-Berry Phase Nonlinear Metasurfaces. *Nat. Commun*. **2021**, *12*, 30.

(4) Wiesendanger, R. Nanoscale Magnetic Skyrmions in Metallic Films and Multilayers: A New Twist for Spintronics. *Nat. Rev. Mater*. **2016**, *1*, 16044.

(5) Matthews, M. R.; Anderson, B. P.; Haljan, P. C.; Hall, D. S.; Wieman, C. E.; Cornell, E. A. Vortices in a Bose-Einstein Condensate. *Phys. Rev. Lett*. **1999**, *83*, 4.

(6) Hsieh, D.; Qian, D.; Wray, L.; Xia, Y.; Hor, Y. S.; Cava, R. J.; Hasan, M. Z. A Topological Dirac Insulator in a Quantum Spin Hall Phase. *Nature* **2008**, *452*, 970-974.

(7) Hsieh, D.; Xia, Y.; Wray, L.; Qian, D.; Pal, A.; Dil, J. H.; Osterwalder, J.; Meier, F.; Bihlmayer, G.; Kane, C. L.; Hor, Y. S.; Cava, R. J.; Hasan, M. Z. Observation of Unconventional Quantum Spin Textures in Topological Insulators. *Science* **2009**, *323*, 919–922.

(8) Moore, J. E. The Birth of Topological Insulators. *Nature* **2010**, *464*, 194–198.

(9) Lu, L.; Joannopoulos, J. D.; Soljačić, M. Topological Photonics. *Nat. Photonics* **2014**, *8*, 821–829.

(10) Allen, L.; Beijersbergen, M. W.; Spreeuw, R. J. C.; Woerdman, J. P. Orbital Angular Momentum of Light and the Transformation of Laguerre-Gaussian Laser Modes. *Phys. Rev. A* **1992**, *45*, 8185–8189.

(11) Simpson, N. B.; Dholakia, K.; Allen, L.; Padgett, M. J. Mechanical Equivalence of Spin and Orbital Angular Momentum of Light: An Optical Spanner. *Opt. Lett*. **1997**, *22*, 3.

(12) Tsesses, S.; Ostrovsky, E.; Cohen, K.; Gjonaj, B.; Lindner, N. H.; Bartal, G. Optical Skyrmion Lattice in Evanescent Electromagnetic Fields. *Science* **2018**, *361*, 993–996.

(13) Fert, A.; Cros, V.; Sampaio, J. Skyrmions on the Track. Nat. *Nanotechnol*. **2013**, *8*, 152–156.

(14) Sampaio, J.; Cros, V.; Rohart, S.; Thiaville, A.; Fert, A. Nucleation, Stability and Current-Induced Motion of Isolated Magnetic Skyrmions in Nanostructures. *Nat. Nanotechnol.* **2013**, *8*, 839–844.

(15) Yu, G.; Upadhyaya, P.; Shao, Q.; Wu, H.; Yin, G.; Li, X.; He, C.; Jiang, W.; Han, X.; Amiri, P. K.; Wang, K. L. Room-Temperature Skyrmion Shift Device for Memory Application. *Nano Lett.* **2017**, *17*, 261–268.

(16) Mühlbauer, S.; Binz, B.; Jonietz, F.; Pfleiderer, C.; Rosch, A.; Neubauer, A.; Georgii, R.; Böni, P. Skyrmion Lattice in a Chiral Magnet. *Science* **2009**, *323*, 915–919.

(17) Al Khawaja, U.; Stoof, H. Skyrmions in a Ferromagnetic Bose–Einstein Condensate. *Nature* **2001**, *411*, 918–920.

(18) Fukuda, J.; Žumer, S. Quasi-Two-Dimensional Skyrmion Lattices in a Chiral Nematic Liquid Crystal. *Nat. Commun.* **2011**, *2*, 246.

(19) Sondhi, S. L.; Karlhede, A.; Kivelson, S. A.; Rezayi, E. H. Skyrmions and the Crossover from the Integer to Fractional Quantum Hall Effect at Small Zeeman Energies. *Phys. Rev. B* **1993**, *47*, 16419–16426.

(20) Davis, T. J.; Janoschka, D.; Dreher, P.; Frank, B.; Meyer zu Heringdorf, F.-J.; Giessen, H. Ultrafast Vector Imaging of Plasmonic Skyrmion Dynamics with Deep Subwavelength Resolution. *Science* **2020**, *368*, eaba6415.

(21) Gao, S.; Speirits, F. C.; Castellucci, F.; Franke-Arnold, S.; Barnett, S. M.; Götte, J. B. Paraxial Skyrmionic Beams. *Phys. Rev. A* **2020**, *102*, 053513.

(22) Gutiérrez-Cuevas, R.; Pisanty, E. Optical Polarization Skyrmionic Fields in Free Space. *J. Opt.* **2021**, *23*, 024004.

(23) Kuratsuji, H.; Tsuchida, S. Evolution of the Stokes Parameters, Polarization Singularities, and Optical Skyrmion. *Phys. Rev. A* **2021**, *103*, 023514.

(24) Lin, W.; Ota, Y.; Arakawa, Y.; Iwamoto, S. Microcavity-Based Generation of Full Poincaré Beams with Arbitrary Skyrmion Numbers. *Phys. Rev. Res.* **2021**, *3*, 023055.

(25) Yang, J.; Zheng, X.; Wang, J.; Pan, Y.; Zhang, A.; Cui, T. J.; Vandenbosch, G. A. E. Symmetry-Protected Spoof Localized Surface Plasmonic Skyrmion. *Laser Photonics Rev.* **2022**, *16*, 2200007.

(26) Shen, Y.; Martínez, E. C.; Rosales-Guzmán, C. Generation of Optical Skyrmions with Tunable Topological Textures. *ACS Photonics* **2022**, *9*, 296–303.

(27) Shen, Y.; Hou, Y.; Papasimakis, N.; Zheludev, N. I. Supertoroidal Light Pulses as Electromagnetic Skyrmions Propagating in Free Space. *Nat. Commun.* **2021**, *12*, 5891.

(28) Deng, Z.-L.; Shi, T.; Krasnok, A.; Li, X.; Alù, A. Observation of Localized Magnetic Plasmon Skyrmions. *Nat. Commun.* **2022**, *13*, 8.

(29) Dai, Y.; Zhou, Z.; Ghosh, A.; Kapoor, K.; Dąbrowski, M.; Kubo, A.; Huang, C.-B.; Petek, H. Ultrafast Microscopy of a Twisted Plasmonic Spin Skyrmion. *Appl. Phy. Rev.* **2022**, *9*, 011420.

(30) Du, L.; Yang, A.; Zayats, A. V.; Yuan, X. Deep-Subwavelength Features of Photonic Skyrmions in a Confined Electromagnetic Field with Orbital Angular Momentum. *Nat. Phys.* **2019**, *15*, 650–654.

(31) Shi, P.; Du, L.; Yuan, X. Strong Spin–Orbit Interaction of Photonic Skyrmions at the General Optical Interface. *Nanophotonics* **2020**, *9*, 4619–4628.

(32) Zhang, Q.; Xie, Z.; Du, L.; Shi, P.; Yuan, X. Bloch-Type Photonic Skyrmions in Optical Chiral Multilayers. *Phys. Rev. Res.* **2021**, *3*, 023109.

(33) Shi, P.; Du, L.; Li, M.; Yuan, X. Symmetry-Protected Photonic Chiral Spin Textures by Spin–Orbit Coupling. *Laser Photonics Rev.* **2021**, *15*, 2000554.

(34) Lei, X.; Du, L.; Yuan, X.; Zayats, A. V. Optical Spin–Orbit Coupling in the Presence of Magnetization: Photonic Skyrmion Interaction with Magnetic Domains. *Nanophotonics* **2021**, *10*, 3667–3675.

(35) Shi, P.; Du, L.; Yuan, X. Spin Photonics: From Transverse Spin to Photonic Skyrmions. *Nanophotonics* **2021**, *10*, 3927–3943.

(36) Lei, X.; Yang, A.; Shi, P.; Xie, Z.; Du, L.; Zayats, A. V.; Yuan, X. Photonic Spin Lattices: Symmetry Constraints for Skyrmion and Meron Topologies. *Phys. Rev. Lett.* **2021**, *127*, 237403.

(37) Zhang, Q.; Xie, Z.; Shi, P.; Yang, H.; He, H.; Du, L.; Yuan, X. Optical Topological Lattices of Bloch-Type Skyrmion and Meron Topologies. *Photon. Res.* **2022**, *10,* 947.

(38) Fei, Z.; Rodin, A. S.; Andreev, G. O.; Bao, W.; McLeod, A. S.; Wagner, M.; Zhang, L. M.; Zhao, Z.; Thiemens, M.; Dominguez, G.; Fogler, M. M.; Neto, A. H. C.; Lau, C. N.; Keilmann, F.; Basov, D. N. Gate-Tuning of Graphene Plasmons Revealed by Infrared Nano-Imaging. *Nature* **2012**, *487*, 82–85.

(39) Chen, J.; Badioli, M.; Alonso-González, P.; Thongrattanasiri, S.; Huth, F.; Osmond, J.; Spasenović, M.; Centeno, A.; Pesquera, A.; Godignon, P.; Zurutuza Elorza, A.; Camara, N.; de Abajo, F. J. G.; Hillenbrand, R.; Koppens, F. H. L. Optical Nano-Imaging of Gate-Tunable Graphene Plasmons. *Nature* **2012**, *487*, 77–81.

(40) Zheng, Z. B.; Li, J. T.; Ma, T.; Fang, H. L.; Ren, W. C.; Chen, J.; She, J. C.; Zhang, Y.; Liu, F.; Chen, H.-J.; Deng, S.-Z.; Xu, N.-S. Tailoring of Electromagnetic Field Localizations by Two-Dimensional Graphene Nanostructures. *Light Sci. Appl.* **2017**, *6*, e17057.

BRIEFS (Word Style "BH_Briefs"). If you are submitting your paper to a journal that requires a brief, provide a one-sentence synopsis for inclusion in the Table of Contents.

# Supporting Information

# Néel-type optical skyrmions inherited from evanescent electromagnetic fields with rotational symmetry

*Bo Tian[1,†], Jingyao Jiang[1,†], Ningsheng Xu[1,2], Zebo Zheng[1], Ximiao Wang[1], Shaojing Liu[1], Wuchao Huang[1], Tian Jiang[3,*], Huanjun Chen[1,*], Shaozhi Deng[1,*]*

[1]State Key Laboratory of Optoelectronic Materials and Technologies, Guangdong Province Key Laboratory of Display Material and Technology, School of Electronics and Information Technology, Sun Yat-sen University, Guangzhou 510275, China.

[2]The Frontier Institute of Chip and System, Fudan University, Shanghai 200433, China.

[3]Institute for Quantum Information Science and Technology, College of Science, National University of Defense Technology, Changsha 410073, China.

*chenhj8@mail.sysu.edu.cn; *tjiang@nudt.edu.cn; *stsdsz@mail.sysu.edu.cn;

**Note S1: The definition of the angle $\varphi$ between the vector of the Néel type skyrmion and 2D plane supporting the skyrmion**

The Néel type skyrmions in 3D space are shown in Figure S1a. It can be found that the angle $\varphi$ between the vector of the Néel type skyrmion and 2D plane supporting the skyrmion can be divided into four situations $\varphi_1$, $\varphi_2$, $\varphi_3$, and $\varphi_4$ (Figure S1b). For a simpler description, we specify that when the vector points toward the positive z direction, the angle $\varphi$ is positive, while it is negative when the vector points toward the negative z direction. Therefore, in Figure S1b, the angles $\varphi_1$ and $\varphi_4$ are positive, while the angles $\varphi_2$ and $\varphi_3$ are negative. The angle at a specific point can be expressed as:

$$\varphi = \begin{cases} \arctan(e_z/e_r), & e_r \geq 0 \\ -\arctan(e_z/e_r), & e_r < 0 \end{cases} \tag{S1}$$

where $e_z$ and $e_r$ are the z-component and in-plane component of the vector of the Néel type skyrmion, respetively.

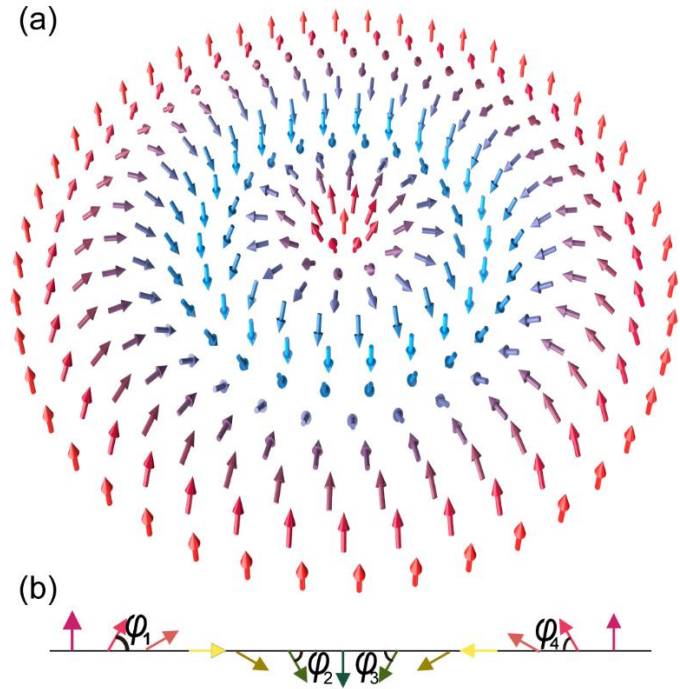

**Figure S1.** Néel type skyrmions. (a) Distribution of the vector fields in a Néel type skyrmion with rotational symmetry. (b) Distribution of the vector fields along a specific radial direction, showing a cycling from the up state ($\varphi = 90°$) at the center of the skyrmion to down state ($\varphi = -90°$) and then back to up state along the radial direction. The angle between the vector and the 2D plane where the skyrmion is located can be divided into four situations as shown in (b).

**Note S2: The skyrmion number of the skyrmion with rotational symmetry**

The topological invariant $S$ of a skyrmion is defined in Cartesian coordinate system as:

$$S = \frac{1}{4\pi} \iint_D \vec{e} \cdot \left( \partial_x \vec{e} \times \partial_x \vec{e} \right) dxdy \tag{S2}$$

where $D$ is the integration area, $\vec{e}$ is the normalized three-dimensional vector. When the angular component of the $\vec{e}$ in the cylindrical coordinate system ($r$, $\theta$, $z$) is 0, the vector $\vec{e}$ can be expressed as:

$$\vec{e} = \cos\varphi \hat{i}_r + \sin\varphi \hat{i}_z \tag{S3}$$

Where $\varphi$ is a function of the coordinates $r$, indicating the angle between the vector $\vec{e}$ and the plane where $D$ is located. Because the relationships between the unit direction vector $\hat{i}_r$ and $\hat{i}_\theta$ in the cylindrical coordinate system and the unit vector $\hat{i}_x$ and $\hat{i}_y$ in the Cartesian coordinate system are

$$\hat{i}_r = \frac{x}{\sqrt{x^2+y^2}} \hat{i}_x + \frac{y}{\sqrt{x^2+y^2}} \hat{i}_y \tag{S4a}$$

$$\hat{i}_\theta = \frac{-y}{\sqrt{x^2+y^2}} \hat{i}_x + \frac{x}{\sqrt{x^2+y^2}} \hat{i}_y \tag{S4b}$$

Therefore, the derivative of the unit direction vector $\hat{i}_r$ and $\hat{i}_\theta$ with respect to the independent variables $x$ and $y$ are

$$\frac{\partial \hat{i}_r}{\partial x} = \frac{y^2}{r^3}\hat{i}_x - \frac{xy}{r^3}\hat{i}_y = -\frac{y}{r^2}\hat{i}_\theta \tag{S5a}$$

$$\frac{\partial \hat{i}_r}{\partial y} = -\frac{xy}{r^3}\hat{i}_x + \frac{x^2}{r^3}\hat{i}_y = \frac{x}{r^2}\hat{i}_\theta \tag{S5b}$$

Then the derivative of the normalized unit vector $\vec{e}$ defined in equation (S3) with respect to the independent variable $x$ and $y$ are

$$\frac{\partial \vec{e}}{\partial x} = \cos\varphi \frac{\partial \varphi}{\partial x}\hat{i}_z - \sin\varphi \frac{\partial \varphi}{\partial x}\hat{i}_r - \frac{y}{r^2}\cos\varphi \hat{i}_\theta \text{ (S6a)} \quad \frac{\partial \vec{e}}{\partial y} = \cos\varphi \frac{\partial \varphi}{\partial y}\hat{i}_z - \sin\varphi \frac{\partial \varphi}{\partial y}\hat{i}_r + \frac{x}{r^2}\cos\varphi \hat{i}_\theta \tag{S6b}$$

Then the form of the integrand in equation (S2) can be expressed as

$$\vec{e}\cdot\left(\frac{\partial \vec{e}}{\partial x}\times\frac{\partial \vec{e}}{\partial y}\right) = -\frac{\cos\varphi}{r^2}\left(y\frac{\partial \varphi}{\partial y} + x\frac{\partial \varphi}{\partial x}\right) = -\frac{\cos\varphi}{r}\frac{\partial \varphi}{\partial r} \tag{S7}$$

Therefore, for the vector distribution as shown in equation (S3), the topological invariance of skyrmion can be transformed as

$$S = -\frac{1}{4\pi}\iint_D \frac{\cos\varphi}{r}\frac{\partial \varphi}{\partial r}r dr d\theta = -\frac{1}{2}\int_D \cos\varphi \frac{\partial \varphi}{\partial r}dr \tag{S8}$$

According to Equation (S8), $S$ is always equal to 1 ($-1$) as long as the angle $\varphi$ changes from $\pi/2$ ($-\pi/2$) to $-\pi/2$ ($\pi/2$).

**Note S3: The relationship between the different components of electric fields of TM electromagnetic waves in a cylindrical coordinate system**

When both the current density and charge density are zero, Maxwell's equations can be expressed as:

$$\nabla \times \vec{E} = \frac{1}{r}\begin{bmatrix} \vec{e}_r & r\vec{e}_\theta & \vec{e}_z \\ \frac{\partial}{\partial r} & \frac{\partial}{\partial \theta} & \frac{\partial}{\partial z} \\ E_r & rE_\theta & E_z \end{bmatrix} = i\omega\mu_0 \vec{H} \tag{S9}$$

and:

$$\nabla \times \vec{H} = \frac{1}{r}\begin{bmatrix} \vec{e}_r & r\vec{e}_\theta & \vec{e}_z \\ \frac{\partial}{\partial r} & \frac{\partial}{\partial \theta} & \frac{\partial}{\partial z} \\ H_r & rH_\theta & H_z \end{bmatrix} = -i\omega\varepsilon_0 \vec{E} \tag{S10}$$

By expanding Equation (S9) and (S10), one can lead to the following relations between different components of the electric field and magnetic field strengths:

$$\frac{1}{r}\frac{\partial E_z}{\partial \theta} - \frac{\partial E_\theta}{\partial z} = i\omega\mu_0 H_r \tag{S11a}$$

$$\frac{\partial E_r}{\partial z} - \frac{\partial E_z}{\partial r} = i\omega\mu_0 H_\theta \tag{S11b}$$

$$\frac{1}{r}\frac{\partial (rE_\theta)}{\partial r} - \frac{1}{r}\frac{\partial E_r}{\partial \theta} = i\omega\mu_0 H_z \tag{S11c}$$

and:

$$\frac{1}{r}\frac{\partial H_z}{\partial \theta} - \frac{\partial H_\theta}{\partial z} = -i\omega\varepsilon_0 E_r \tag{S12a}$$

$$\frac{\partial H_r}{\partial z} - \frac{\partial H_z}{\partial r} = -i\omega\varepsilon_0 E_\theta \tag{S12b}$$

$$\frac{1}{r}\frac{\partial(rH_\theta)}{\partial r} - \frac{1}{r}\frac{\partial H_r}{\partial \theta} = -i\omega\varepsilon_0 E_z \tag{S12c}$$

For a TM ($H_z = 0$) electromagnetic waves, the dependence of the electric field on the coordinate $z$ and $t$ is $\exp(ik_z'z - i\omega t)$. By plugging Equation (S12a) into (S11b), one can obtain:

$$E_r = \frac{ik_z'}{k_s^2}\frac{\partial E_z}{\partial r} \tag{S13}$$

By plugging the equation (S12b) to (S11a), one can obtain:

$$E_\theta = \frac{ik_z'}{rk_s^2}\frac{\partial E_z}{\partial \theta} \tag{S14}$$

where $k_s$ is the in-plane wave vector, and satisfies $k_s^2 + k_z'^2 = k_0^2$. $k_0 = \omega\sqrt{\mu_0\varepsilon_0}$ is the wave vector in vacuum. When $k_z' = ik_z$ is an imaginary number, the electric field is an evanescent wave in the $z$-direction and the dependence of the electric field on the coordinate $z$ is $\exp(-k_z z)$. Equation (S13) and (S14) then become:

$$\begin{cases} E_r = -\dfrac{k_z}{k_s^2}\dfrac{\partial E_z}{\partial r} \\ E_\theta = -\dfrac{k_z}{rk_s^2}\dfrac{\partial E_z}{\partial \theta} \end{cases} \tag{S15}$$

where the wave vector satisfies $k_s^2 - k_z^2 = k_0^2$.

**Note S4: Fourier transform of rotationally symmetric functions**

The Fourier transform (FT) of a binary functions $f(x,y)$ is defined as:

$$F(k_x, k_y) = \frac{1}{2\pi} \int_{-\infty}^{\infty} \int_{-\infty}^{\infty} f(x,y) e^{-i(k_x x + k_y y)} dxdy \tag{S16}$$

where $(x, y)$ and $(k_x, k_y)$ are the independent variables in real space and momentum space, respectively. The relations between the different variables in Cartesian coordinate system and polar coordinate system are:

$$x = r\cos\theta, \ y = r\sin\theta, \ k_x = \rho\cos\beta, \ k_y = \rho\sin\beta \tag{S17}$$

where $r$ and $\rho$ are the independent polar variables in real space and momentum space, respectively. Therefore, the FT (S16) can be expressed as:

$$F(\rho\cos\beta, \rho\sin\beta) = \frac{1}{2\pi} \int_0^{2\pi} \int_0^{\infty} f(r\cos\theta, r\sin\theta) e^{-i\rho r\cos(\theta-\beta)} rdrd\theta \tag{S18}$$

When the function $f$ is independent on $\theta$, Equation (S18) can be expressed as:

$$F(\rho\cos\beta, \rho\sin\beta) = \frac{1}{2\pi} \int_0^{2\pi} \int_0^{\infty} f(r) e^{-i\rho r\cos(\theta-\beta)} rdrd\theta \tag{S19}$$

Due to the existence of definite integral:

$$\int_0^{2\pi} e^{-i\rho r\cos(\theta-\beta)} d\theta = 2\pi J_0(\rho r) \tag{S20}$$

Equation (S19) can be simplified as:

$$F(\rho\cos\beta, \rho\sin\beta) = \int_0^{\infty} rf(r) J_0(\rho r) dr \tag{S21}$$

Because the right-hand side of the Equation (S21) is not a function of $\beta$, the left-hand side of the equation should not be a function of angle $\beta$ either. Therefore, Equation (S21) can be expressed as:

$$F(\rho) = \int_0^\infty r f(r) J_0(\rho r) dr \tag{S22}$$

Similarly, the expressions of the inverse FT is:

$$f(r) = \int_0^\infty \rho F(\rho) J_0(\rho r) d\rho \tag{S23}$$

According to the above analysis, when a function $f$ is rotationally symmetric, its FT $F$ must be rotationally symmetric, and vice versa. Equation (S22) and (S23) are 2D FT of rotationally symmetric functions, also known as the zero-order Hankel transform.

For the rotationally symmetric function $C\delta(\rho-k_s)$ illustrated in the lower left corner of Figure 2a in the main text, its inverse FT can be found by equation (S23). Then field intensity is:

$$E_z(r) = \int_0^\infty \rho C(\rho - k_s) J_0(\rho r) d\rho = C k_s J_0(k_s r) \tag{S24}$$

Since the function decays exponentially in the z-direction, the entire distribution of the electric field intensity can be expressed as

$$E_z(r,z) = C k_s J_0(k_s r) e^{-k_z z} \tag{S25}$$

**Note S5: Methodology on processing the experimental near-field optical signals and calculating the different electric field components**

The near-field amplitude (Figure S1a) and phase (Figure S1b) signals are obtained by a s-SNOM system. For rotationally symmetric functions, the entire 2D structure can be obtained by rotating the radial function for 360° around the origin. Therefore, in order to better prove that the TM evanescent waves with rotational symmetry are skyrmions, we extracted the experimental data in the radial direction in the experimental diagram (the positions marked by the dashed lines in Figure S1a and S1b) and the extracted results are shown in Figure S3c, where the values of the amplitude $|E_z|$ and phase are shown in the left and right axes, respectively. The real part of the electric field is obtained by multiplying amplitude and the cosine of the phase, which is $E_{z,\text{exp}}$ shown in Figure S1d. In the experimental data, high-frequency noises and the background light are inevitable. We eliminated these noises by filtering on FT of the raw data. In Figure S1d, we show the experimental data $E_{z,\text{fliter}}$ obtained by eliminating high frequency noise and $E_{z,\text{background}}$ obtained by eliminating the combined effect of high frequency noise and background light. By comparing $E_{z,\text{exp}}$ and $E_{z,\text{fliter}}$, it can be found that the results after high-frequency filtering only makes the curve smooth, and the other features are exactly the same as those from the experimental data. The frequency corresponding to the background light is of zero-frequency and the intensity of the background light is independent on the spatial coordinate, therefore, $E_{z,\text{background}}$ and $E_{z,\text{fliter}}$ only differ by a constant. Since the s-SNOM only detects the axial component of the electric field, it is necessary to calculate the in-plane component of the electric field by Equation (S15). After normalizing the electric field, the $e_z$ and $e_r$ can be obtained, as shown in Figures 3b and 3c in the main text.

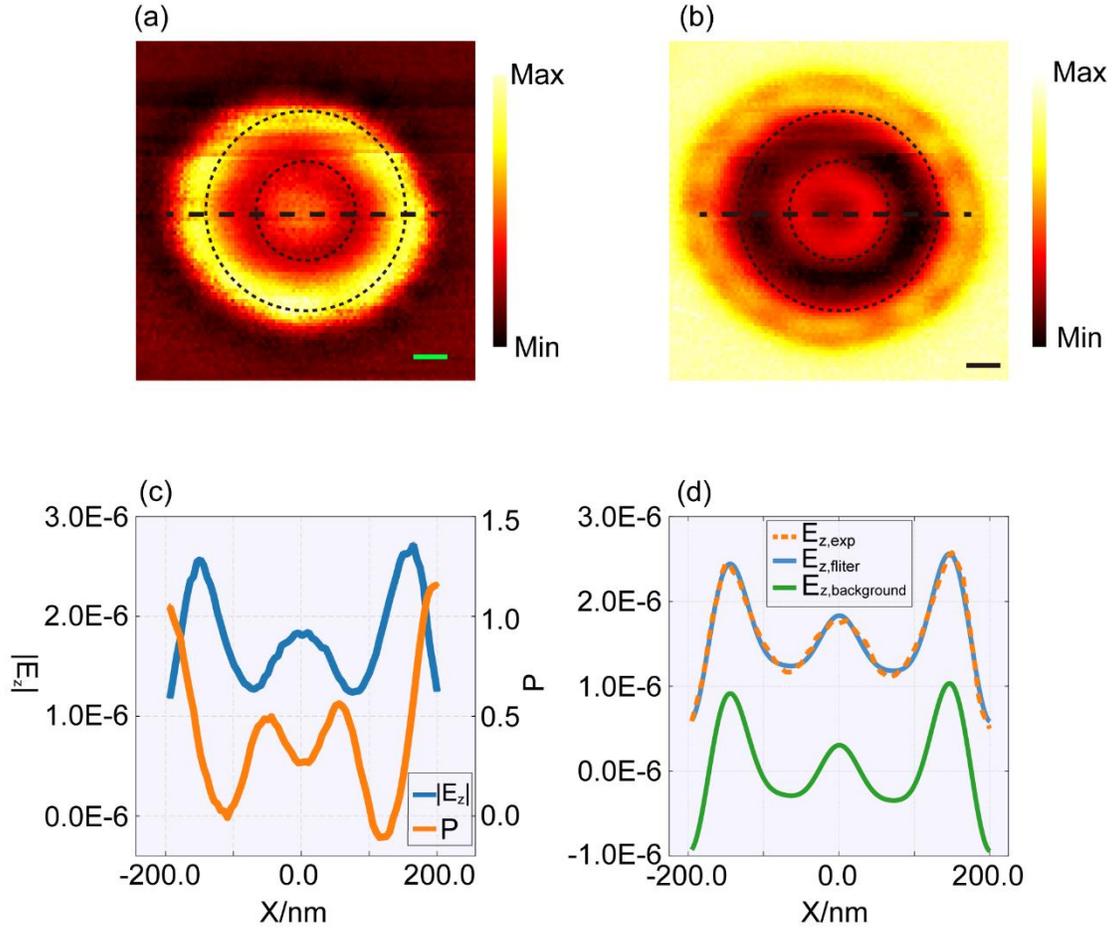

**Figure S2.** Optical skyrmions observed on a graphene circular disk. (a, b) The near-field magnitude (a) and phase (b) of *z*-component of the electric field observed on the monolayer graphene. The solid black line marks the raw data selected for data processing. (c) The amplitude $|E_z|$ and phase P extracted at the dashed line marked in (a) and (b), where the $|E_z|$ and P correspond to the left and right axes, respectively. (d) The real part, $E_{z,\text{exp}}$, of the electric field. $E_{z,\text{exp}}$ is obtained by multiplying the amplitude and the cosine of the phase. The $E_{z,\text{fliter}}$ is the experimental results obtained by filtering the high-frequency component in the FT spectrum of $E_{z,\text{exp}}$. $E_{z,\text{background}}$ is obtained by eliminating the background noise from $E_{z,\text{fliter}}$. The scale bars in (a) and (b) are 50 nm.

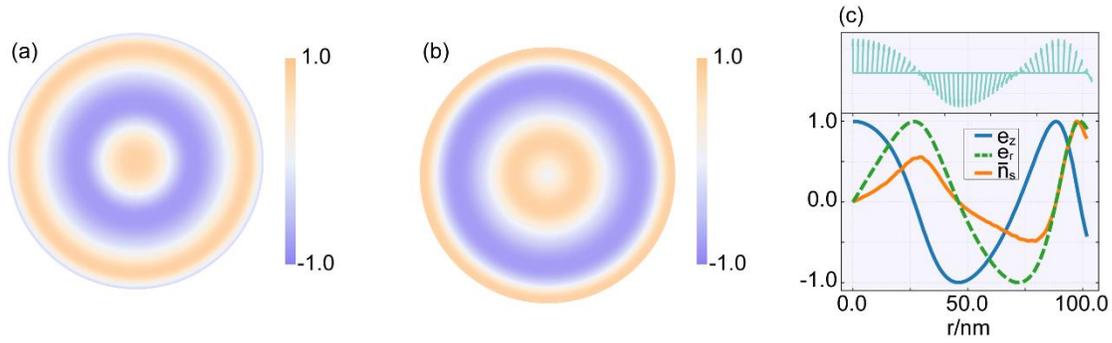

**Figure S3.** Néel-type optical skyrmions formed by surface plasmon polaritons in a monolayer graphene. (a, b) Distributions of the normalized electric field component $e_z$ (a) and $e_r$ (b) above the graphene surface. (c) Radial variations of the $e_z$, $e_r$, and $\bar{n}_s$ along a specific in-plane direction. Upper panel in (c): radial evolvement of the electric field vector. A $z$-polarized electric dipole is employed as the excitation, which is placed 100 nm above the monolayer graphene. The electric field signals were obtained on the plane 50 nm above the sample surface. The excitation wavelength is 10.00 μm.

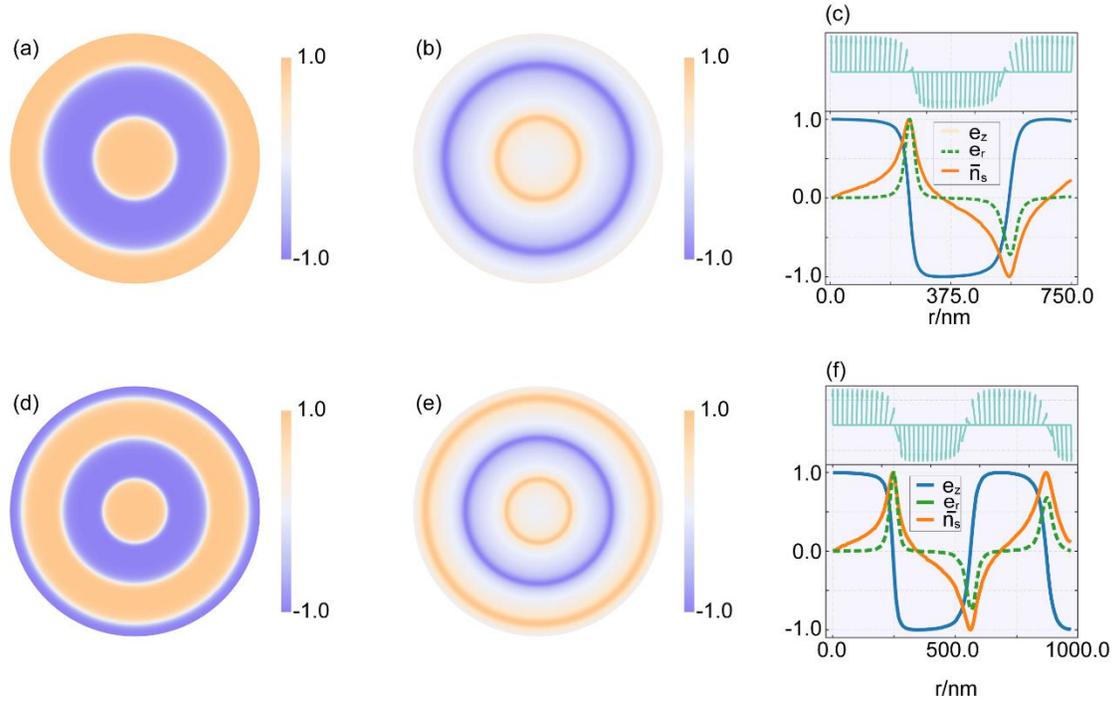

**Figure S4.** Néel-type optical skyrmions formed by surface plasmon polaritons in a gold flak. (a, b, d, e) Distributions of the normalized electric field component $e_z$ (a, d) and $e_r$ (b, e) above the gold flake. (c, f) Radial variations of the $e_z$, $e_r$, and $\bar{n}_s$ along a specific in-plane direction. Upper panels in (c) and (f): radial evolvements of the electric field vector. A $z$-polarized electric dipole is employed as the excitation, which is placed 100 nm above the gold flake. The electric field signals were obtained on the plane 50 nm above the sample surface. The excitation wavelengths are 532 nm (a–c) and 633 nm (d–e), respectively. The thickness of the gold flake is 200 nm.

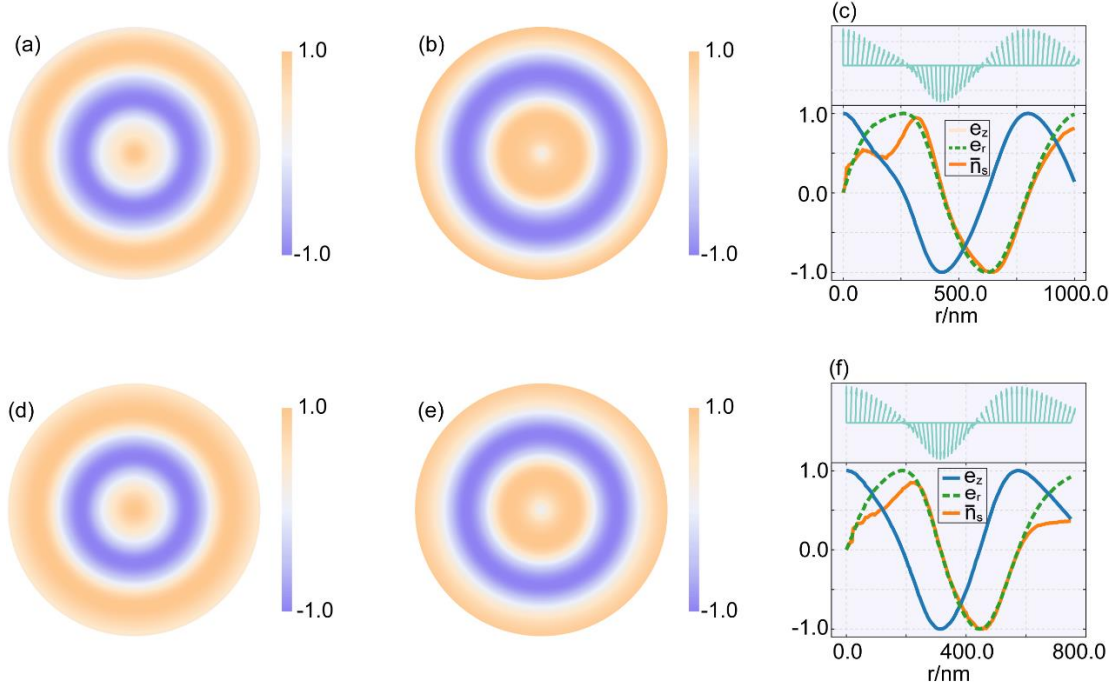

**Figure S5.** Néel-type optical skyrmions formed by phonon polaritons in an hBN flak. (a, b, d, e) Distributions of the normalized electric field component $e_z$ (a, d) and $e_r$ (b, e) above the gold flake. (c, f) Radial variations of the $e_z$, $e_r$, and $\bar{n}_s$ along a specific in-plane direction. Upper panels in (c) and (f): radial evolvements of the electric field vector. A $z$-polarized electric dipole is employed as the excitation, which is placed 100 nm above the hBN flake. The electric field signals were obtained on the plane 50 nm above the sample surface. The excitation wavelengths are 12.66 μm (a−c) and 12.50 μm (d−e), respectively. The thickness of the hBN flake is 100 nm.

**Table S1.** The coordinates of the zeros of $e_r$ and $\bar{n}_s$, and the extreme points of $e_z$ in Figures 1−3 in the main text.

| Radial coordinate | | $e_r$ | $\bar{n}_s$ | $e_z$ |
|---|---|---|---|---|
| **Figure1:** $k_s r$ | | 3.832 | 3.832 | 3.832 |
| | | 7.0156 | 7.0156 | 7.0156 |
| | | 10.173 | 10.173 | 10.173 |
| | | 13.324 | 13.324 | 13.324 |
| **Figure 2:** $r$ (nm) | | 52.4 | 52.4 | 52.4 |
| | | 97.5 | 97.5 | 97.5 |
| **Figure 3:** $r$ (nm) | | 72 | 72 | 72 |
| | | 146 | 146 | 146 |

**Table S2.** The values of the topological invariants calculated in Figure 2 and Figure 3 in the main text.

| | topological invariant S | |
|---|---|---|
| **Figure 2** | 0.995 | −0.994 |
| **Figure 3** | 1.007 | −1.001 |